\documentclass[12pt]{article}

\usepackage{latexsym}
\newif\ifams\amsfalse
\amstrue \ifams
 \usepackage{amsfonts}
\fi

\newcommand{\bea}{\begin{eqnarray}}
\newcommand{\eea}{\end{eqnarray}}
\newcommand{\be}{\begin{equation}}
\newcommand{\ee}{\end{equation}}

\def\cta{\tilde{S}^{A}}

\def\ctb{\tilde{S}^{B}}

\ifams
 \def\RR{\mathbb R} \else
 \def\RR{\mbox{\rm$\mbox{I}\!\mbox{R}$}}
\fi

\begin{document}

\title{
\begin{flushright}
\begin{small}
 hep-th/9906127\\
 EFI-99-29 \\
 KUL-TF-99/23\\
 June 1999 \\
\end{small}
\end{flushright}
\vspace{1.cm}
The Gravitational Action in\\
Asymptotically AdS and Flat Spacetimes}
\author{
 Per Kraus$^{1}$, Finn Larsen$^{1}$ and Ruud Siebelink$^{1,2}$
 \thanks{Post-doctoraal Onderzoeker FWO, Belgium}\\
 \small \vspace{- 3mm} $^1$ the Enrico Fermi Institute  \\
 \small \vspace{- 3mm} University of Chicago\\
 \small 5640 S. Ellis Ave., Chicago, IL-60637, USA\\
 \small \vspace{- 3mm} $^2$ Instituut voor Theoretische Fysica\\
 \small \vspace{- 3mm} Katholieke Universiteit Leuven\\
 \small \vspace{- 3mm} Celestijnenlaan 200D\\
 \small B-3001 Leuven, Belgium \\
 \small E-mail: {\tt pkraus,flarsen,siebelin@theory.uchicago.edu} }
\date{}
\maketitle

\begin{abstract}
The divergences of the gravitational action are analyzed for
spacetimes that are asymptotically anti-de Sitter and
asymptotically flat. The gravitational action is rendered finite
using a local counterterm prescription, and the relation of this
method to the traditional reference spacetime is discussed. For
AdS, an iterative procedure is devised that determines the
counterterms efficiently. For asymptotically flat space, we use a 
different method to derive counterterms which are sufficient to 
remove divergences in most cases.
\end{abstract}

\newpage

\section{Introduction}
\label{sec:intro}
The concepts of action and energy-momentum  play
central roles in theories with gravity, but they are surprisingly
difficult to define (see, {\em e.g.}, \cite{wald}), and they are often
laborious to compute. A well known obstacle to the straightforward
definition of the gravitational action in a non-compact space is
that the sum of the Einstein-Hilbert and Gibbons-Hawking terms
diverges.  There is a standard remedy for this calamity: first
regulate the divergence by restricting the spacetime to the
interior of some bounding surface, then subtract the (similarly
infinite) action of some reference spacetime with the same
boundary geometry~\cite{Gibbons:1977ue}.  For an appropriately
chosen reference spacetime, the resulting action will be finite as
the boundary is taken to infinity.  The energy-momentum tensor of
the spacetime is then related to the variation of the total action
with respect to the boundary metric~\cite{Brown:1993br}.

The above procedure suffers from a number of important drawbacks.  On a
conceptual level, it is not satisfying since it relies on the
introduction
of a spacetime which is auxiliary to the problem.  It is sometimes said
that this merely corresponds to defining the overall zero of energy, but
in fact the procedure also affects {\em relative} energies,
because different reference spacetimes are needed for different
boundary geometries.  A more
glaring defect is that the procedure is generally ill-defined, since it
is
not possible to embed an arbitrary boundary geometry in the reference
spacetime.  One is forced to resort to an approximate embedding, and
this
often leads to confusion and ambiguity;  good examples of this are
Taub-NUT and Taub-bolt
spacetimes
~\cite{Hunter:1998qe,Chamblin:1998pz,Hawking:1998jf,Hawking:1998ct,Taylor-Robinson:1998fc}.

Recently, a preferable alternative procedure has been
proposed~\cite{Balasubramanian:1999re}.
For a manifold with boundary, the only way to modify the gravitational 
action without disturbing the equations of motion or the symmetries 
is to add a coordinate invariant functional of the intrinsic boundary 
geometry.  By choosing a functional --- which we refer to as a 
counterterm --- which cancels the divergences, one arrives at well 
defined expressions for the action and energy-momentum of the spacetime.  
The procedure is satisfying since it is intrinsic to the spacetime of 
interest, and unambiguous once the counterterm is specified.

The new prescription is particularly elegant in the case of
asymptotically AdS spacetimes~\cite{Balasubramanian:1999re}. In
these cases the structure of the divergences is such that they can
be fully removed by adding a finite polynomial in the boundary
curvature and its derivatives. Moreover, the counterterm method is
precisely analogous to the standard removal of divergences in
quantum field theory by adding finite polynomials in the fields;
indeed, the AdS/CFT correspondence~\cite{Aharony:1999ti} asserts
that they are the very same.  The equivalence between the
gravitational action and that of a CFT on the boundary is
illustrated by the agreement between trace
anomalies~\cite{Henningson:1998gx,Hyun:1998vg,Nojiri:1998dh,Muck:1999nf} and
Casimir energies~\cite{Balasubramanian:1999re,Myers:1999ps}
obtained from the two descriptions (see~\cite{Chalmers:1999gc} for
a different appearance of divergences in the AdS/CFT
correspondence).

Counterterms for low dimensional AdS spacetimes were obtained
in~\cite{Balasubramanian:1999re,Emparan:1999pm}. Our focus in the
first half of this paper is to develop an algorithm for generating
counter\-terms for arbitrary dimensions. We begin by analyzing the
structure of divergences in more detail. An important tool is the
interplay between the bulk and boundary geometries, as embodied in
the Gauss-Codazzi equations. This formulation of the problem leads
to an iterative process that generates the counterterms as an
expansion in the radius of the anti-de Sitter space, denoted by
$\ell$:
 \be
 {\tilde S} = {1 \over\ell} [{\tilde S}^{(0)} + \ell^2
 {\tilde S}^{(1)} +\cdots ]
 \ee
The coefficients $\tilde{S}^{(n)}$ are increasingly complicated
polynomials of the boundary curvature tensor and its derivatives.
The iterative process is manifestly covariant and quite efficient;
we compute the four leading orders explicitly, and show that our
algorithm defines a series of local counterterms to all orders.
The trace anomalies of the boundary theory play a prominent role
in the discussion; indeed, there is a close connection to the
anomaly computation of~\cite{Henningson:1998gx}.

Crucial to the success of the counterterm prescription is that the
divergences are universal, so that a single choice of counterterms
suffices to render finite the action of all asymptotically AdS
solutions\footnote{The action will sometimes have logarithmic
divergences which remain uncancelled by the counterterms. However,
these divergences are well understood physically as arising from
the trace anomaly.}. The finite terms are
non-universal, though, and so the counterterm subtraction leaves a
finite remainder in general. We use the AdS-Schwarzschild
spacetimes to illustrate these properties of the action.    An
important feature of the counterterm prescription is that it
provides unambiguous results even for nontrivial boundary
topologies, such as
Taub-NUT-AdS~\cite{Mann:1999pc,Emparan:1999pm}.

It is of obvious interest to apply the counterterm method to
spacetimes that are asymptotically flat, rather than
asymptotically AdS. One approach is to try to determine these by
solving the flat space versions of the Gauss-Codazzi equations.
However, the resulting equation is highly non-linear and we have
had no progress with this strategy. An alternative strategy is to
define flat space as the limit where the AdS curvature
$\ell\to\infty$.  But to take the limit one first has to determine
the counterterms to all orders in $\ell$, which seems
prohibitively difficult.  We pursue a third approach, which can be
understood as a refinement of the reference spacetime
prescription.  We derive the action of a particular spacetime
which asymptotes to the solutions of interest, and write the
result in terms of intrinsic invariants of the boundary.  The
resulting counterterm action is  then expected to share the
divergences of spacetimes which look sufficiently like the chosen
spacetime near infinity. Unlike in the standard reference
spacetime prescription, once we have obtained the counterterm we
can forget about the reference spacetime altogether.

Recently Lau~\cite{Lau:1999dp} and Mann~\cite{Mann:1999pc}
proposed the remarkably simple counter\-term for spaces that are
asymptotically $AdS_{4}$:
 \be
 8\pi G {\tilde S} = - {2\over\ell}\int\! d^3 x
 \,\sqrt{-\gamma}\sqrt{1 + {\ell^2\over 2}R}~,
 \ee
which has a smooth limit as $\ell\to\infty$. Mann further showed
that, in many explicit examples, this removes all divergences and
gives a finite part that agrees with the reference spacetime
procedure~\cite{Mann:1999pc,Mann:1999bt}. By following the
strategy of the previous paragraph we derive the $d$-dimensional
generalization of the Lau-Mann formula.  However, the assumptions
of the derivation are quite strong, and there are simple examples
where divergences are {\it not} removed. We give a more general
counterterm that removes the divergences for asymptotically flat
space in more cases, though not in general. Our examples suggest
that a counterterm capable of removing the divergences from
arbitrary  asymptotically flat spacetimes would be quite
complicated.  However, we stress that such an expression is not
needed under normal circumstances --- the counterterms we define
provide well defined actions for the most common class of
spacetimes.  A more general result would only be needed if one
wished to consider spacetimes which deviate strongly from these.

All of the counterterms that we derive will be coordinate invariant,
intrinsic to the boundary, and local.   
The property of locality is not {\em a priori} mandatory, since
adding non-local intrinsic counterterms would not disturb the equations of
motion and so cannot be excluded on such grounds. As we will
describe below, for asymptotically AdS spaces the divergences are always
local polynomials in the boundary fields and their derivatives, as  was to be
expected given the AdS/CFT correspondence.  In the case
of asymptotically flat spaces it is less clear what to expect,
since we do not know whether the putative holographically
dual theories will have only local divergences. Still, in the simple
asymptotically flat examples we consider below, it suffices to use
local countertems only in order to remove the divergences.

The flat space limit of AdS space has recently been discussed in
the context of the complete, dynamical string
theory~\cite{Polchinski:1999ry,Susskind:1998vk}. It was argued
that, in the appropriate limit, the AdS/CFT  correspondence
constitutes a suitable starting point for non-perturbative
M-theory in flat space. The divergences in the gravitational
action which we study give nontrivial information concerning a
possible holographic description of flat space. The correct
understanding of the flat space counterterm may ultimately be
interlinked with these far-reaching perspectives.

The paper is organized as follows. In section 2 we develop our
algorithm for generating AdS counterterms, and give examples. We
turn to flat space in section 3, and give two examples of
counterterms.  The counterterms are seen to lead to the standard
results for the actions of black hole spacetimes, provided that
appropriate coordinates are chosen. We show in an example how more
general coordinates may lead to ambiguities, whose nature we
explain.

\section{Counterterms and the Gauss-Codazzi equation}

We write the standard action for the gravitational field as
\footnote{This fixes our conventions for the Riemann curvature to
be $R_{\mu\nu\lambda}{}^{\sigma}= - 2 \partial_{[\mu}
\Gamma_{\nu]\lambda}{}^{\sigma} + 2 \Gamma_{\lambda [\mu}
{}^{\rho} \Gamma_{\nu ]\rho}{}^{\sigma}$, where the
antisymmetrization is defined with strength one, i.e. $[\mu \nu ]
= \frac{1}{2} (\mu\nu - \nu\mu)$. Also $R_{\mu\nu} =
R_{\mu\lambda\nu} {}^\lambda$. With these conventions spheres have
a {\it positive} scalar curvature. The cosmological constant is
written as $\Lambda=-d(d-1)/2\ell^2$; in this notation pure
$AdS_{d+1}$ has radius $\ell$.}:
\be
  S= \frac{1}{16\pi G}\int_M d^{d+1}x \sqrt{- g}
  \left( R + \frac{d(d-1)}{\ell^2}\right)
  -\frac{1}{8\pi G}\int_{\partial M} d^d x \sqrt{-\gamma} ~ \Theta~.
\label{action}
\ee
Variation of this action with respect to the geometry of the boundary
$\partial M$ gives the energy-momentum tensor~\cite{Brown:1993br}:
\be
\Pi^{ab} = \Theta^{ab}-\Theta\gamma^{ab}~,
\label{emtensor}
\ee
where $\gamma_{ab}$ is the metric of the boundary. Concrete
computations show that in most spacetimes both the action integral
(\ref{action}) and the energy-momentum tensor (\ref{emtensor}) actually
diverge as the boundary $\partial M$ is taken to infinity. We therefore
think of these as the unrenormalized quantities.

The divergences must be cancelled in order to achieve physically
meaningful
expressions; {\it i.e.} some counterterm action:
\be
{\tilde S} = \frac{1}{8\pi G}\int d^d x \sqrt{-\gamma} ~ {\tilde{\cal
L}}~,
\ee
must be added, along with the corresponding counterterm energy-momentum
tensor:
\be
 \tilde{\Pi}^{ab} = {2\over\sqrt{-\gamma}}~
{\delta\over\delta\gamma_{ab}}
\int d^d x \sqrt{-\gamma}~\tilde{\cal L}~.
 \label{defpct}
\ee The counterterms by definition contain the divergent part of
the corresponding unrenormalized quantities, but finite terms may
depend on the details of the renormalization.

Now, recall that the Gibbons-Hawking boundary term in
(\ref{action}) has been determined precisely such that the
combined action satisfies a well-defined variational principle,
giving the correct bulk equation of motion. The counterterm will
ruin this property unless it is a function of the boundary
geometry only. Additionally, suppose the counterterm is an
analytical function of the boundary geometry, and expand it as a
power series in the metric and its derivatives. Dimensional
analysis shows that in $AdS_{d+1}$ only terms of order $n < d/2$
contribute to the divergent part of the action. (By terms of order $n$
we mean terms containing $2n$ derivatives.) Therefore one may truncate
the series at this order and obtain a finite
polynomial~\cite{Balasubramanian:1999re}. This agrees with the
expectations from the interpretation of the divergences in terms
of a dual boundary theory that obeys the usual axioms of quantum field
theory, including locality. In most of this paper we will treat
the dimension $d$ as a free parameter which can be made very
large and so think of the counterterm as a
$d$-dependent power series with an arbitrarily high number of terms.
Of course, when restricting the attention to a particular value
of $d$, the general result should be truncated.

\subsection{The Counterterm Generating Algorithm}
The structure of divergences is tightly constrained
by the Gauss-Codazzi equations. These are covariant expressions of the
bulk Einstein tensor $G_{\mu\nu} = R_{\mu\nu} -\frac{1}{2} g_{\mu\nu} R$
in terms of the boundary Einstein tensor $G_{ab}(\gamma)$ (which only
depends on the induced metric $\gamma_{ab}$) and the extrinsic
curvature $\Theta_{ab}$ (which characterizes the embedding of the
boundary surface into the bulk geometry).
After using (\ref{emtensor}) they read~\cite{wald}:
\bea
 G_{ab} &=& G_{ab}(\gamma) + {\hat n}^\mu \nabla_\mu \Pi_{ab}
 - \frac{1}{2} \gamma_{ab} \left( \frac{\Pi^2}{d-1} - \Pi_{cd} \Pi^{cd}
 \right)
 + \frac{1}{d-1} \Pi_{ab} \Pi~,
 \label{GC1}\\
 G_{a\mu} {\hat n}^\mu &=& - \nabla^b \Pi_{ba} \label{GC2}~, \\
 G_{\mu\nu} {\hat n}^\mu {\hat n}^\nu &=& \frac{1}{2} \left(
 \frac{1}{d-1} \Pi^2 - \Pi_{ab} \Pi^{ab} - R(\gamma) \right) ~,
 \label{GC3}
 \eea
where ${\hat n}^\mu$ is an outward pointing unit vector normal to the
boundary. We will always consider
solutions of the bulk equations of motion so:
 \bea
 G_{ab} &=& \frac{1}{2} \frac{d(d-1)}{\ell^2} \gamma_{ab}~, \nonumber\\
 G_{a\mu} {\hat n}^\mu &=& 0~, \nonumber \\
 G_{\mu\nu} {\hat n}^\mu {\hat n}^\nu &=& \frac{1}{2}
\frac{d(d-1)}{\ell^2}~,
 \eea
determines the left hand side of the Gauss-Codazzi
equations.

In principle, one could solve the Gauss-Codazzi equations
(\ref{GC1})-(\ref{GC3}) for the unrenormalized energy-momentum
tensor $\Pi_{ab}$, and then identify its divergent part with $-
{\tilde\Pi}_{ab}$. However, this strategy is rather complicated
due to the presence of the normal derivatives in (\ref{GC1}). The
appearance of these normal derivatives expresses the intuitive
fact that, to determine the solution throughout, both the boundary
values and their derivatives are needed. However, the counterterm
should be determined independently of data that is extrinsic to
the boundary, such as the normal derivative.

Now, there exists a set of coordinates for which the bulk Einstein
equations in $AdS_{d+1}$
(which are equivalent to the Gauss-Codazzi equations)
can be solved in a perturbative fashion~\cite{fefgra}. Explicit
computations in this coordinate system show that the
{\it divergent part} of the normal derivatives can be expressed in
terms of the {\it intrinsic} boundary data~\cite{fefgra}. We
implement this observation covariantly, as follows. We impose the
constraint
equation (\ref{GC3}):
\be
 \frac{1}{d-1} {\tilde\Pi}^2 - {\tilde\Pi}_{ab}
 {\tilde\Pi}^{ab} = \frac{d(d-1)}{\ell^2} + R~,
 \label{defining1}
 \ee
and further insist that the counterterm energy-momentum tensor must
derive from a counterterm action, which is itself intrinsic to the
boundary:
\be
 \tilde{\Pi}^{ab} = {2\over\sqrt{-\gamma}}~
 {\delta\over\delta\gamma_{ab}}
 \int d^d x \sqrt{-\gamma}~\tilde{\cal L}~.
 \label{defining2}
 \ee
As we will show, the conditions (\ref{defining1}) and (\ref{defining2})
fully
determine the counterterm. The form of (\ref{defining2}) ensures
that the counterterm energy-momentum is conserved, which in turn
implies (\ref{GC2}). It is important to stress that the remaining
Gauss-Codazzi equations (\ref{GC1}) are also satisfied: they can be
viewed as expressions for the normal derivatives specified implicitly
in our construction. We note that the normal derivatives thus determined
do not in general vanish.

There is another implicit definition of the normal derivative
which is important in the AdS/CFT correspondence: this is the
requirement of {\it regularity} in the bulk of
spacetime~\cite{Witten:1998qj}. For Euclidean $AdS_{d+1}$ with a
boundary sufficiently close to the round sphere $S^d$ it is known
that, for given intrinsic boundary data, there exists a unique
solution to the Gauss-Codazzi equations which is regular in the
interior of $AdS$~\cite{gralee}. This is not in general the
solution we consider. That our solution may become singular when
expanded to all orders is of no concern because, for a specific
value of the boundary dimension $d$, we always truncate to a
finite number of terms.

We are now prepared to describe an algorithm that determines the
counterterm as an expansion in the parameter $\ell$. The leading
order term scale as $\ell^{-1}$ and terms at a given order
$\ell^{2n-1}$ with $n\geq 0$ are denoted by
${\tilde\Pi}_{ab}^{(n)}$ and ${\tilde{\cal L}}^{(n)}$. The
starting point is to note that the curvature term in
(\ref{defining1}) can be neglected to the leading order in $\ell$,
so that the metric is the only tensor characterizing the boundary
geometry to the leading order. The ${\tilde\Pi}_{ab}^{\rm (0)}$
are therefore proportional to the metric, with the overall
numerical factor determined by (\ref{defining1}). This gives:
\be
{\tilde\Pi}_{ab}^{\rm (0)} = - \frac{d-1}{\ell} \gamma_{ab}~.
\label{leadpi}
\ee
The sign was determined using positivity conditions on the
energy-momentum tensor.

Higher order counterterms are now given by induction. Assuming
that ${\tilde\Pi}_{ab}$ is known up to and including order $n-1$,
the following three steps determine ${\tilde\Pi}_{ab}^{(n)}$:
\paragraph{step 1:}
Insert the known terms in (\ref{defining1}); the resulting
expression is a linear equation with the trace ${\tilde\Pi}^{(n)}$
as the only unknown.
\paragraph{step 2:}
With the trace $\tilde{\Pi}^{(n)}$ in hand, integrate
(\ref{defining2}) and find $\tilde{\cal L}^{(n)}$. This step is
purely algebraic, as discussed in the following subsection.
\paragraph{step 3:}
Finally, take the functional derivative of $\tilde{\cal L}^{(n)}$
with respect to $\gamma_{ab}$, and so find the full tensor
$\tilde{\Pi}_{ab}^{(n)}$ from (\ref{defining2}).

\vspace{.5cm} The fact that ${\tilde\Pi}_{ab}^{\rm (0)}$ is
proportional to the metric $\gamma_{ab}$ is crucial to make step 1
possible. We stress that higher orders of ${\tilde\Pi}_{ab}$ in
general will depend also on other tensor structures.

\subsection{Some Comments on Weyl Rescalings}
The integration in step 2 is interesting and deserves comment. It
is related to the behavior of the various terms under the local
Weyl variations which transform the metric as:
\be
 \delta_W \gamma_{ab} = \sigma \gamma_{ab}~,
 \label{weyl}
\ee where $\sigma$ is an arbitrary function. Consider the
counterterm action at the $n$th order and note that dimensional
analysis gives the behavior under a {\it global} Weyl rescaling.
The result of a {\it local} Weyl variation can therefore be
written in the form:
 \be
 \delta_W \int d^d x \sqrt{- \gamma} \, {\tilde{\cal L}}^{(n)}
 =
 \int d^d x \sqrt{- \gamma} \, \sigma \, \left( \frac{d-2n}{2}
 {\tilde{\cal L}}^{(n)} + \nabla_a X^{a(n)} \right)~,
 \label{delL}
 \ee
where $X^{a(n)}$ is some unspecified expression (involving $2n+1$
derivatives).\footnote{It is important that the Lagrangian
${\tilde{\cal L}}^{(n)}$ is assumed to be local, otherwise
(\ref{delL}) need not be true. For example, $\delta_W \int d^d x
\sqrt{- \gamma} \, R \frac{1}{\Box} R = \int d^d x \sqrt{- \gamma}
\, \sigma \, \left( \frac{d-2}{2} R \frac{1}{\Box} R - 2(d-1)R
\right)$.} However, it follows from (\ref{defining2}) that:
 \be
 \delta_W \int d^d x \sqrt{- \gamma} \, \tilde{\cal L}^{(n)} =
 \frac{1}{2} \int d^d x
 \sqrt{- \gamma}~\sigma~\tilde{\Pi}^{(n)}~,
 \label{weylL}
 \ee
so:
 \be
 (d-2n) {\tilde{\cal L}}^{(n)} = \tilde{\Pi}^{(n)}~,
 \label{intid}
 \ee
up to a total derivative term. Now, recall that counterterm
Lagrangians are in fact only defined up to total derivatives; a
total derivative term can be added without changing the action. We
can therefore freely choose a scheme where (\ref{intid}) is {\it
exact}, without need for total derivatives. The practical
significance of this identity is that it renders the integration
in step 2 almost trivial. We also note that:
 \be
 \delta_W \int d^d x \sqrt{- \gamma} \, {\tilde\Pi}^{(n)}  =
 \frac{d-2n}{2} \int d^d x \sqrt{-\gamma} ~\sigma ~\Pi^{(n)}~,
 \label{weyltensor}
 \ee
so that $\sqrt{- \gamma}~\Pi^{(n)}$ transforms as a conformal
density with Weyl weight $\frac{1}{2}(d-2n)$, up to total
derivatives~\cite{Bonora:1986cq,Deser:1993yx}. This constrains the
form of the counterterms.

In even dimensions it is clear that (\ref{intid}) prevents
${\tilde\Pi}^{(d/2)}$ from being obtained as the variation of any
local action. This is the origin of trace anomalies. For a given
even dimension $d$ the trace ${\tilde\Pi}^{(d/2)}$ is therefore
identified with the trace anomaly of the dual boundary theory.
This result for the anomaly agrees with that
of~\cite{Henningson:1998gx}, as may be verified by looking at the
explicit expressions given below.

\subsection{Explicit Computations of Counterterms}
At this point we evaluate the counterterms explicitly,
to the first few orders.

The leading order was given in (\ref{leadpi}). For completeness we
give its trace and the corresponding counterterm, computed using
(\ref{intid}):
 \bea
 {\tilde\Pi}^{(0)} &=&  - \frac{d(d-1)}{\ell}~, \nonumber\\
 {\tilde{\cal L}}^{(0)} &=& - \frac{d-1}{\ell}~.
 \eea
At the first non-trivial order we insert this in (\ref{defining1})
and find:
 \be
 {\tilde\Pi}^{(1)} = - {\ell\over 2}R~.
 \ee
Now (\ref{intid}) gives:
\be
{\tilde {\cal L}}^{(1)} = - {\ell\over 2(d-2)}R~,
\ee
and the variation (\ref{defining2}) yields:
 \be
 {\tilde\Pi}^{(1)}_{ab} = {\ell\over d-2}\left( R_{ab}-
 {1\over2}\gamma_{ab}R
 \right)~.
 \ee
We have used the algorithm to
generate a few more orders in the expansion, finding the trace of
the energy-momentum tensor:
 \bea
 \tilde{\Pi}
 &=& - \frac{d(d-1)}{\ell} - \frac{l}{2} R
 - \frac{\ell^3}{2(d-2)^2} \left( R_{ab} R^{ab} - \frac{d}{4(d-1)} R^2
 \right)\nonumber \\
 && + \frac{\ell^5}{(d-2)^3(d-4)} \left\{
 \frac{3d+2}{4(d-1)} R R_{ab} R^{ab}
 - \frac{d(d+2)}{16 (d-1)^2} R^3
 - 2 R^{ab} R_{acbd} R^{cd} \right. \nonumber \\
 && + \left. \frac{d-2}{2(d-1)} R^{ab} \nabla_a \nabla_b  R
 - R^{ab} \Box R_{ab}
 + \frac{1}{2(d-1)} R \Box R \right\} + \cdots~,
 \label{trace}
 \eea
the counterterm Lagrangian:
 \bea
 {\tilde{\cal L}}
 &=& - \frac{d-1}{\ell} - \frac{\ell}{2(d-2)} R
 - \frac{\ell^3}{2(d-2)^2(d-4)} \left( R_{ab} R^{ab} -
 \frac{d}{4(d-1)} R^2 \right)\nonumber \\
 && + \frac{\ell^5}{(d-2)^3(d-4)(d-6)} \left\{
 \frac{3d+2}{4(d-1)} R R_{ab} R^{ab}
 - \frac{d(d+2)}{16 (d-1)^2} R^3
 - 2 R^{ab} R_{acbd} R^{cd} \right. \nonumber \\
 &&  \left.
 + \frac{d-2}{2(d-1)} R^{ab} \nabla_a \nabla_b  R
 - R^{ab} \Box R_{ab}
 + \frac{1}{2(d-1)} R \Box R \right\} + \cdots~,
\label{counter}
\eea
and the full energy-momentum tensor:
\bea
 \tilde\Pi_{ab}
 &=& - \frac{d-1}{\ell} \gamma_{ab}
 + \frac{\ell}{d-2} \left( R_{ab} - \frac{1}{2} \gamma_{ab} R \right)
 \nonumber \\
 && + \frac{\ell^3}{(d-2)^2(d-4)} \left\{
 - \frac{1}{2} \gamma_{ab} \left( R_{cd} R^{cd} - \frac{d}{4(d-1)} R^2
 \right) - \frac{d}{2(d-1)} R R_{ab} \right. \nonumber \\
 && \left.  + 2 R^{cd} R_{cadb}
 - \frac{d-2}{2(d-1)} \nabla_a \nabla_b  R
 + \Box R_{ab}
 - \frac{1}{2(d-1)} \gamma_{ab} \Box R \right\} + \cdots~.
 \label{full}
\eea
The most laborious step is to find the full energy momentum-tensor
from the counter-term. Accordingly, we have resisted carrying out
this computation to the fourth order.

The first three orders of (\ref{trace}) agree with the results
previously
deduced from explicit
examples~\cite{Balasubramanian:1999re,Emparan:1999pm}. All four 
terms may be obtained from the results in~\cite{Henningson:1998gx}
(see in particular the Appendix of the second reference.)

\subsection{Explicit examples in AdS}
We now consider a few simple examples that illustrate the general
results.

\subsubsection{Euclidean $AdS_{d+1}$ with a $S^d$ boundary}

Consider pure $AdS_{d+1}$ with the metric:
 \be
 ds^2 = \frac{dr^2}{1+ r^2/\ell^2} + r^2 d\Omega^2_{d}~.
 \ee
Using the definition (\ref{emtensor}) of the unrenormalized
energy-momentum tensor, one can straightforwardly compute the
density:
 \bea
 \sqrt{\gamma}~\Pi &=& \frac{d(d-1)}{\ell}~r_0^d~
 \sqrt{1 + \frac{\ell^2}{r_0^2}} ~\sqrt{g_{\Omega_d}} \nonumber\\
 &=& \frac{d(d-1)}{\ell}~
 \left( r_0^d  + \frac{1}{2}~l^2 r_0^{d-2} - \frac{1}{8}~l^4 r_0^{d-4}
 + \cdots \right)  \sqrt{g_{\Omega_d}}
 ~.
\label{divex}
\eea
It clearly diverges for $r_0 \rightarrow \infty$, so one needs to
add counterterms in order to cancel the terms of order $n<d/2$.
The Ricci scalar on the boundary is expressed in terms of the position
$r_0$ of the boundary through:
\be
 R =\frac{d(d-1)}{r_0^2}~.
\label{rint}
\ee
Inserting this in the formula for the counterterm $\tilde\Pi$ given
in (\ref{trace}), and exploiting that the space is maximally symmetric,
we
find terms that are precisely equal to (\ref{divex}), with the opposite
sign. Thus the divergences cancel.

However, as emphasized above, the counterterm (\ref{trace}) should
be truncated so that only the divergences cancel. For even
dimension this implies that the finite term in  (\ref{divex}) is
{\it not} subtracted. This residual term is the anomaly. That
there is a genuine obstruction that precludes the cancellation of
this term is seen by inspecting the would-be counterterm action
(\ref{counter}) at order $d=n/2$. After evaluation on $S^d$ this
term remains ill-defined due to the $1/(d-2n)$ factor.

For odd-dimensional boundaries there are no anomalies and the
renormalized energy-momentum tensor vanishes. However, this does
not imply that the renormalized action vanishes. To see this,
consider the contribution to the action from the bulk part of
(\ref{action}). This term has the form of a radial integral
$\int^{r_0}_0 d r$ leading to a term at $r=r_0$, which is in fact
cancelled by the counterterms, but also a term at $r=0$ which does
not get subtracted off. Interestingly, the resulting uncancelled
action implies negative entropy for the case of $AdS_4$ with an
$S^3$ boundary~\cite{Emparan:1999pm}.

\subsubsection{The AdS-Schwarzschild solution}
Next, we consider the AdS-Schwarzschild metric:
\be
 ds^2 = 
 - \left( 1+ \frac{r^2}{\ell^2} - \frac{\mu}{r^{d-2}} \right) dt^2
+ \frac{dr^2}{\left( 1+ \frac{r^2}{\ell^2} -
 \frac{\mu}{r^{d-2}} \right)}
 + r^2 d\Omega_{d-1}^2~,
\ee which for large values of $r$ asymptotes to a pure $AdS_{d+1}$
solution with a $\RR\times S^{d-1}$ boundary. The corresponding
unrenormalized energy-momentum tensor (\ref{emtensor}) can be
expanded as:
 \bea
 \Pi^{\hat a \hat b} &=& \frac{\gamma^{\hat a \hat b}}{\ell} \left\{
(d-1)
 + \frac{d-3}{2}~\frac{\ell^2}{r_0^2}
 - \frac{d-5}{8}~\frac{\ell^4}{r_0^4}+ \cdots
 + \frac{\mu \ell^2}{2 r_0^d} + {\cal O} (r_0^{-(d+1)})\right\}~,
 \nonumber \\
 \Pi^{tt} &=& \frac{\gamma^{tt}}{\ell} \left\{
 (d-1) + \frac{d-1}{2}~\frac{\ell^2}{r_0^2}
 - \frac{d-1}{8}~\frac{\ell^4}{r_0^4}+ \cdots
 - (d-1) \frac{\mu \ell^2}{2 r_0^d}  + {\cal O} (r_0^{-(d+1)})\right\}~,
 \nonumber \\
 \Pi &=&  \frac{(d-1)}{\ell} \left\{
 d + \frac{d-2}{2}~\frac{\ell^2}{r_0^2}
 - \frac{d-4}{8}~\frac{\ell^4}{r_0^4}
 + \cdots + {\cal O} (r_0^{-(d+1)})\right\}~,
 \label{expr}
 \eea
where the index $\hat a$ labels the $S^{d-1}$ directions. In order
to see which of the terms in (\ref{expr}) correspond to divergent,
finite or vanishing physical quantities in the $r_0\to\infty$
limit, one must convert the above expressions to the appropriate
proper densities. One finds that the proper scaling of the
quantities in (\ref{expr}) is determined by multiplying the
formulae inside the brackets by $r^d_0$; thus the ${\cal O}
(r_0^{-(d+1)})$ terms vanish in the limit.

It is very important that the divergent terms depend on the
boundary position $r_0$ only. This ensures that these terms are
{\it intrinsic} to the boundary, as they should be. (That the
divergent terms are intrinsic can be made manifest by expressing $r_0$
in terms of the Ricci scalar $R$, using $R=(d-1)(d-2)/r_0^2$.) The
mass parameter $\mu$ is an {\it extrinsic} quantity from the
boundary point of view, so it is important that the
$\mu$-dependence appears only at the finite level. Because of its
extrinsic nature, the $\mu$-dependence can never be subtracted
off.

Generically (\ref{intid}) forces a $1/(d-2n)$ divergence in the
counterterm action at order $n=d/2$, rendering the corresponding
subtraction impossible. The present example is special because the
trace $\Pi$ contains a $(d-2n)$ factor at each order $n$ so that a
finite counterterm at order $n=d/2$ is viable. The physical
significance of this possibility is that we can choose to include
the order $d/2$ counterterm as well, such that the entire
$\mu$-independent part of the energy-momentum tensor is cancelled.
It was shown in~\cite{Balasubramanian:1999re} that the
$\mu$-independent finite part of the energy-momentum tensor can be
identified with the Casimir energy in the dual conformal field
theory. The ability to cancel this part of the energy-momentum
tensor is equivalent to the option of choosing a renormalization scheme
where the Casimir energy of $\RR\times S^{d-1}$ is set to zero.

\section{Counterterms for Asymptotically Flat Space}
We now turn to defining the gravitational action in asymptotically
flat space (AFS). At first glance, AFS may seem like just a
special case of AdS, since it can be obtained by taking the limit
$\ell\rightarrow\infty$. This analogy leads us to consider
(\ref{defining1}), which in the limit reads:
\be
 \frac{1}{d-1} {\tilde\Pi}^2 - {\tilde\Pi}_{ab}
{\tilde\Pi}^{ab} = R~, \ee and we should further impose
(\ref{defining2}) on the solution. The problem is now nonlinear
and does not appear to allow a perturbative expansion; a direct
computation is therefore impractical. We might instead try to
apply the limit $\ell\to\infty$ to our perturbative expansion of
the counterterm action. However, the large $\ell$ limit can only
be taken after summing the infinite series, which is clearly a
difficult task. It is also doubtful whether the sum exists, for
the following reasons: As we mentioned before, the counterterm
action (\ref{counter}) implicitly defines a bulk solution which in
general need not be regular. Therefore, if the sum did exist, it
would generically assign some finite action to singular solutions,
which seems unphysical. Most glaringly, for even $d$ the
coefficients of individual terms diverge.

We will take an alternative approach to define the AFS counterterms.
We will start with some particular solution, work out its action, and
then express the result in terms of intrinsic invariants of the
boundary.
The counterterm action is then defined as minus this expression.
To the extent that divergences are universal, this counterterm action
will remove the divergences of solutions which asymptote to the
particular
solution used in the construction. The counterterm is not uniquely
defined, since choosing different solutions or different curvature
invariants will yield inequivalent
results. As we will discuss, this does not appear to be a drawback of
the
procedure. An added benefit is that this method exhibits the relation
between the counterterm method and the reference spacetime approach
clearly.

\subsection{A Counterterm for AFS}
We first consider the most common class of metrics, those having
boundary topology $S^{d-1}\times \RR$.\footnote{The case of $d=2$
requires special considerations, so we assume $d>2$. It is
interesting that the $d=2$ case is exactly solvable for all
boundary geometries.} In this case we can work out a simple closed
form counterterm for AdS with finite $\ell$, and then take the
flat space limit $\ell \rightarrow \infty$. To do so we consider
$AdS_{d+1}$ in global coordinates:
\be
ds^2 = -(1+r^2/\ell^2)dt^2 + {dr^2 \over 1+r^2/\ell^2} + r^2
d\Omega_{d-1}^2~. \label{globalads} \ee To evaluate the
gravitational action (\ref{action}) we use that
$R-2\Lambda=-2d/\ell^2$ for pure $AdS_{d+1}$, and also the general
expression:
\be
S_{GH} = - {1\over 8\pi G}\int d^d x\sqrt{-\gamma}~\Theta
=  {1\over 8\pi G}\int d^d x {\partial\over\partial{\hat
n}}\sqrt{-\gamma}~,
\label{ghcomp}
\ee
where $\hat{n}$ is the unit normal to the boundary. Then the action
becomes:
\be
S_{{\rm bulk}}+ S_{GH} =
 {d-1 \over 8 \pi G \ell} \int \! d^dx\,
 \sqrt{-\gamma}~\sqrt{1 +\ell^2/r^2_0}~.
 \ee
 To express this
in terms of intrinsic invariants we use:
\be
R =  {(d-1)(d-2) \over r^2_0}~,
\ee
for a $d-1$ sphere of radius $r_0$. This leads to the definition:
\be
\cta = -(S_{{\rm bulk}}+ S_{GH}) = -{d-1 \over 8\pi G\ell} \int \!
d^dx\,
\sqrt{-\gamma}~
\sqrt{1+{\ell^2 R \over (d-1)(d-2)}}~.
\label{Slm}
\ee
By definition, $\cta$ will assign vanishing total action to AdS in
global
coordinates.  We further expect it to give {\em finite} action for
solutions which asymptote to (\ref{globalads}). In the flat space
limit we find:
\be
\tilde{{\cal L}}^{A} = - \sqrt{{d-1 \over d-2}}~\sqrt{R}~.
\label{sla} \ee

It is instructive to compare (\ref{Slm}) with the power series
(\ref{counter}). Choosing coordinates $(\hat{a},\tau)$ on
$S^{d-1}\times \RR$ we have:
\be
R_{\hat{a}\hat{b}\hat{c}\hat{d}}= {1\over (d-1)(d-2)}
(g_{\hat{a}\hat{c}}g_{\hat{b}\hat{d}}-
g_{\hat{a} \hat{d}}g_{\hat{b}\hat{c}})R~,\quad
R_{\hat{a} \hat{b}} = {1\over d-1} g_{\hat{a} \hat{b}}R~,
\label{curv}
\ee
as well as $R_{\hat{a}\tau\hat{b}\tau}=R_{\tau\tau}=0$. The power
series (\ref{counter}) becomes:
\be
{\tilde{\cal L}} = - {d-1 \over \ell} - {1 \over 2(d-2)} R +
{\ell^3 \over 8 (d-1)(d-2)^2} R^2 - {\ell^5 \over 16 (d-1)^2
(d-2)^3} R^3 +  \cdots~. \ee This precisely corresponds to the
expansion of $\cta$ in (\ref{Slm}). So it appears that for the
$S^{d-1}\times \RR$ class of boundary geometries we have summed
the series (\ref{counter}). Note that the potentially divergent
factors of $1/(d-4), 1/(d-6),  \ldots$  were cancelled in the
process, leading to a well defined counterterm for arbitrary
$d>2$. Since (\ref{curv}) only relies on $S^{d-1}$ being a
maximally symmetric space, this computation further indicates that
(\ref{Slm}) also cancels the divergences when $S^{d-1}$ is
replaced by a space with constant negative curvature.

For $d=3$ the counterterm (\ref{Slm}) is closely related to an
expression obtained by Lau through different means~\cite{Lau:1999dp}.
It agrees precisely with the counterterm used by
Mann~\cite{Mann:1999bt}.
Indeed, the present reasoning provides a simple derivation of Mann's
proposal, and its generalization to $d$ dimensions.

\subsection{Counterterms and Black Hole Thermodynamics}
Before continuing the main argument we pause to derive a constraint on
the counterterms from Smarr's formula, and also comment on the
(absent) effect of counterterms on the black hole entropy.
A useful reference for the formulas in this section
is~\cite{Myers:1986un}.

Upon continuation to imaginary time, the action of a Euclidean black
hole
solution represents the thermodynamic free energy of the system:
\be
S= \beta M - \mu_J J - S_{ent},
\ee
where we denote entropy by $S_{ent}$, and inverse temperature --- or
equivalently, the periodicity of imaginary time --- by $\beta$.
$J$ represents the angular momentum of the horizon and $\mu_J$ is
its conjugate potential.  Given $S$ as
a function of $\beta$ and $\mu_J$, the mass and angular momentum follow
from:
\be
M = {\partial S \over \partial \beta}, \quad J = -{\partial S \over
\partial\mu_J}~.
\ee
The Bekenstein-Hawking formula $S_{ent} = A/4G$ and the
generalized Smarr formula:
\be
(d-2)M\beta= (d-1)(\mu_J J + {A \over 4G})~,
\ee
give the simple expression for the action:
\be
S =  {\beta M \over d-1}~.
\label{actresult}
\ee
A notable point is that $\mu_J$ and $J$ do not appear explicitly in the
result,  but only via $\beta$.  In sec.~\ref{discussion} we will
compute the action for a general rotating solution and see explicitly
that it only receives contributions from the leading long-range part 
of the metric, which is equivalent to that of a spherically symmetric solution.

As a simple example, we work out the action of the $d+1$ dimensional
Euclidean Schwarzschild solution using the counterterm action
(\ref{sla}):
\be
ds^2 = \left(1-{\mu \over r^{d-2}} \right)d\tau^2 +
{dr^2 \over\left(1-{\mu \over r^{d-2}} \right)} + r^2 d\Omega_{d-1}^2~,
\ee
where $\mu$ is related to the ADM mass by:
 \be
 \mu = { 16 \pi G M \over (d-1) \omega_{d-1}}~.
 \label{mumass}
 \ee
We find (after rotating to Euclidean time):
 \bea
 S &=&   S_{GH} + \cta = -\lim_{r_0 \rightarrow \infty} {\omega_{d-1}
 \beta \over 8 \pi G} \left[ (d-1)r_0^{d-2} - {d \over 2} \mu -
 (d-1)r_0^{d-2}\sqrt{1- {\mu \over r_0^{d-2}}}~\right] \nonumber\\
 &=& {\omega_{d-1} \beta \mu \over 16 \pi G} =
  {\beta M \over d-1}~,
 \label{genac}
 \eea
in agreement with (\ref{actresult}).

As we have discussed and will see again explicitly, one may
sometimes have a choice between several counterterm actions which
all subtract off the infinities but lead to different finite
results. It is important to stress that in the case of black holes
these choices affect only the definition of energy and not the
entropy; the latter is always given by the Bekenstein-Hawking
formula. This can be seen as follows~\cite{Hawking:1980gf} (for
simplicity we consider the spherically symmetric case).  Consider
restricting the imaginary time integration region  in
(\ref{genac}) to some duration $\Delta \tau$.  Let ${\cal W}$ be
the wedge shaped region in the $r-\tau$ plane given by restricting
$\tau$ to the range $\Delta \tau$. The tip of the wedge lies at
the horizon.  To compute the action of ${\cal W}$ we need to
include a boundary term at the tip. Here only the Gibbons-Hawking
term can contribute, since the vanishing of $g_{\tau\tau}$ at the
horizon causes intrinsic invariants to vanish upon integration.
Now, from the Hamilton-Jacobi equation it follows that the action
of a  static ($\tau$ independent) solution obeys:
\be
S = \Delta \tau M~.
\ee
This equation defines the mass of the spacetime, and so is independent
of
choice of counterterms at infinity, although the actual  value of $M$
can be.
Next, consider evaluating the action for the full Euclidean black hole
manifold.  We then set $\Delta \tau = \beta$, {\em and omit the
Gibbons-Hawking
term at the horizon}, since the only boundary is at  $r=r_0$. Therefore
the action is:
\be
S= \beta M - S_{GH}^{hor}~. \ee Finally, a simple computation
yields that $S_{GH}^{hor} = S_{ent} = A/4G$. We have thus
established that the entropy is insensitive to the addition of
boundary terms at infinity.  The entropy is in this sense
renormalization scheme independent.

\subsection{More Divergences and Another Counterterm for AFS}
The counterterm $\cta$ was designed to remove divergences for
boundaries of the form $S^{d-1} \times \RR$, so there is no
guarantee that it will continue to do so for other boundary
topologies.  As a simple example, take flat Euclidean space in
spherical coordinates:
\be
ds^2 = dr^2 + r^2 d\Omega_d^2~.
\ee
Then:
\be
S_{GH} = {d \omega_{d} \over 8 \pi G} r^{d-1}_0, \quad
\cta = -{(d-1) \omega_{d} \over 8 \pi G} \sqrt{d \over d-2}r^{d-1}_0~.
\ee
The sum of the two terms is not finite as the boundary is taken
to infinity $r_0\to\infty$; so the divergences are {\em not removed}
by $\cta$. Note that this problem persists in spacetimes with four
bulk dimensions, {\it i.e.} $d=3$.

By generalizing the derivation that led to $\cta$ we can derive a
counterterm which will remove divergences for a larger class of
spacetimes, including the example just given. Write flat $d+1$
dimensional  space in a form with boundary $S^n \times \RR^{d-n}$:
 \be
 ds^2 = (dt^2 + dx_1^2 + \cdots + dx_{d-1-n}^2) + dr^2 + r^2
 d\Omega_n^2. \label{flatgen}
 \ee
The action is:
 \be
 S_{GH} = {1 \over 8 \pi G} \int \! d^dx \, \sqrt{\gamma}~{n \over r_0},
 \ee
where $\sqrt{\gamma} = r^n_0 \sqrt{g_{\Omega_n}}$~. We wish to
write a counterterm that will remove the divergence at large $r_0$
for arbitrary $n$. To do so it is necessary to use more invariants
than just $R$. We take also $R^{\mu\nu} R_{\mu\nu}$, and use that
for $S^n \times \RR^{d-n}$:
\be
R = {n(n-1) \over r^2_0}, \quad  R_{\mu\nu}R^{\mu\nu} = R^2/n~.
\ee
Then we arrive at:
\be
\ctb = - S_{GH} = -{1 \over 8 \pi G} \int \! d^dx \, \sqrt{\gamma}~
{R^{3/2} \over \sqrt{R^2 -R_{\mu\nu}R^{\mu\nu}}}~.
\label{slb}
\ee
This counterterm reverts to $\cta$ in the special case of $n=d-1$ for
which $ R_{\mu\nu}R^{\mu\nu} = R^2/(d-1)$.  But $\ctb$ is more general
than $\cta$ since it will remove divergences from metrics that are
asymptotic to any of (\ref{flatgen}).

\subsection{Discussion}
\label{discussion} In cases for which both $\cta$ and $\ctb$
remove divergences one can ask whether they will also agree on the
finite part. An interesting feature is that the answer depends on
the coordinates used to describe a given spacetime --- by using a
preferred class of coordinates the two counterterms will agree in
the main cases of interest.

\subsubsection{Counterterms and the ADM prescription}
It is important to verify that under suitable conditions we obtain
results for the mass and angular momentum which agree with the standard
ADM definitions.
To show this, we use that any AFS can be cast in the form:
\begin{eqnarray}
ds^2 &=& -\left(1-{\mu \over r^{d-2}}  +
O\left({1 \over r^{d-1}} \right) \right) dt^2
- \left( {A^{ij} x^i \over r^{d}} +O\left({1 \over r^{d}} \right)
\right)
dx^j dt \nonumber \\
&+&\left[\left(1+{\mu \over r^{d-2}}  +
O\left({1 \over r^{d-1}} \right)\right)\delta_{ij}
+ {e_{ij} \over r^{d-2}} +  O\left({1 \over r^{d-1}} \right) \right]
dx^i dx^j.
\label{ADMcoords}
\end{eqnarray}
The boundary is taken to be at fixed $r^2_0 = x^i x^i$.
$A^{ij}$ is proportional to the angular momentum of the spacetime, and
the symmetric, traceless tensor $e_{ij}$ represents gravitational
radiation.
We will restrict attention to isolated systems, for which $\mu$ and
$A^{ij}$ are constants; the time dependent case requires a more detailed
analysis.

Now, the important point is that upon evaluating the action we
find that the angular momentum and gravitational radiation terms
in the metric make no contribution as $r_0\rightarrow \infty$.
First note that the $r_0$ dependence is such that  only terms in
the action linear in $A^{ij}$ or $e_{ij}$ can potentially
contribute. But $A^{ij}$ cannot appear linearly in the action due
to time reversal symmetry. And $e_{ij}$ can only appear in the
rotationally invariant combinations $e_{ii}$, $e_{ij}x^i x^j$ (sum
on $i,j$); but for $e_{ij}$ traceless, the former vanishes
identically, while the latter vanishes upon integration over the
sphere. Therefore, in the action, only the $\mu$ dependent terms
survive, and the calculation effectively reduces to that for the
Schwarzschild metric, with boundary $S^{d-1}\times \RR$. However,
we already know from our previous computation (\ref{genac}) that
the counterterms $\cta$ and $\ctb$ agree for boundaries of this
form, and indeed by direct calculation we find:
\be
S= S_{GH} + \cta = S_{GH} + \ctb = -{\omega_{d-1} \beta \mu \over 16 \pi
G}~.
\label{ADMres}
\ee
This is the correct result, as we discussed in the derivation of
(\ref{genac}).  Demanding regularity of the Euclidean black hole metric
fixes the relation between $\mu$ and $\beta$.
Then, by differentiating the action with respect to $\beta$ we can read off
the mass of the solutions. In so doing, we find agreement with the ADM
definition (\ref{mumass}).

It is satisfying that our counterterms reproduce the ADM definitions for
the general class of metrics (\ref{ADMcoords}), since these follow
from quite general considerations.  For instance, the ADM mass is the
unique (up to an overall constant) quantity which is conserved,
transforms
as the time component of a Lorentz vector, and is additive for distant
subsystems.  If one is willing to relax one or more of these conditions
then other results are possible, and we will see an explicit example of
this in the following subsection.

\subsubsection{An Example: Spheroidal Coordinates}
To get agreement between $\cta$, $\ctb$, and the standard result for
the action (\ref{ADMres}) it was important to use the preferred
coordinates
(\ref{ADMcoords}).    This fact is illustrated by considering flat space
in non-standard coordinates.  We know that both counterterms assign zero
action to flat space in the form:
\be
ds^2 = -dt^2 + dr^2 + r^2 d\Omega_{d-1}^2~.
\label{spherical}
\ee
Now instead use spheroidal coordinates:
\be
ds^2 = -dt^2
+ {r^2 + a^2 \cos^2{\theta} \over r^2 + a^2} dr^2 + (r^2 +a^2
\cos^2{\theta})
d\theta^2 + \sin^2{\theta} (r^2 +a^2) d\phi^2
+ r^2 \cos^2{\theta} d\Omega_{d-3}^2
\label{spheroidal}
\ee
This is in fact the form of the metric one finds upon setting the mass
parameter to zero in the D-dimensional generalization of the
Kerr metric written in
Boyer-Lindquist coordinates~\cite{Myers:1986un}. In these
coordinates, the action (\ref{ghcomp}) yields:
\be
S_{GH} = {1 \over 8 \pi G} \int \! d^dx \, \left[
d-1 +(d-3) {a^2 \over r^2_0} +
{ a^2\sin^2{\theta}\over r^2_0 + a^2 \cos^2{\theta}} \right]
r^{d-2}_0 \sin{\theta}
\cos^{d-3}{\theta} \sqrt{g_{\Omega_{d-3}}}
\label{sphflat}
\ee
where the measure is $d^d x= dtd\theta d\phi d\Omega_{d-3}$.
The expression in the square bracket forms a kernel with the expansion:
\be
K = (d-1) + \left( d-2 - \cos^2\theta \right){a^2\over r^2_0}
-(\cos^2\theta-\cos^4\theta) {a^4\over r^4_0}+\cdots. \label{Kexp}
\ee After computing the curvature tensors of the boundary geometry
the counterterm (\ref{sla}) can be written as:
\begin{eqnarray}
\cta&=&-{d-1 \over 8 \pi G}  \int \! d^dx \,
{  (r^2 +a^2)^{1/2} \over (r^2 +a^2 \cos^2{\theta})^{1/2}}
\left[ 1
+ {(d-3)\over (d-1)}(1+\cos^2{\theta}){a^2 \over r^2_0} \right.
\nonumber \\
&+&\left.{(d-3)(d-4) \over (d-1)(d-2)} \cos^2{\theta}
{a^4 \over r^4} \right]^{1/2}r^{d-2}_0
\sin{\theta} \cos^{d-3}{\theta} \sqrt{g_{\Omega_{d-3}}}~,
\end{eqnarray}
corresponding to the expansion:
\bea
{\tilde K}^{A} &=& -(d-1) - (d-2-\cos^2\theta){a^2\over r^2_0}
 \\ &~&\quad
-{1\over 2(d-1)}\left( -1 -{2(d^2-3d+1)\over d-2} \cos^2\theta +
(2d-3)\cos^4\theta
\right)
{a^4\over r^4_0}+\cdots
\nonumber
\eea
in the same normalization as (\ref{Kexp}).
The general expression for the alternative counterterm (\ref{slb})
is quite lengthy, so we give just the expansion:
\bea
{\tilde K}^{B} &=& -(d-1) - (d-2-\cos^2\theta){a^2\over r^2_0}
+{1\over (d-1)(d-2)^2}
\left( {2d-5\over 2} + \right.
 \\ &+& \left. {13-22d+18d^2-7d^3+d^4\over d-2}\cos^2\theta
+ {3-14d+10d^2 - 2d^3\over 2}\cos^4\theta \right){a^4\over r^4_0}+\cdots
\nonumber
\eea
These expressions show that both  counterterms correctly capture the
divergences to the first two leading orders. The renormalized action
therefore vanishes as $r_0\to\infty$ for $d<6$, it is finite
for $d=6$, and for $d>6$ the divergences of flat space in spheroidal
coordinates are not cancelled by either counterterm. After integration
over
$\theta$ we find that, for $d=6$, the finite action assigned to flat
space
is nonvanishing, but different for the two counterterms.

On a technical level, note that the metrics (\ref{spherical}),
(\ref{spheroidal}) are equivalent to leading order for large $r$,
with both boundaries asymptoting to $S^{d-1} \times \RR$.  It is
the subleading $a$ dependent terms which lead to the finite terms
in the action.

This discussion was for flat space in spheroidal coordinates. It
is straightforward to compute the expression analogous to
(\ref{sphflat}) for the Kerr black hole in $D$
dimensions~\cite{Myers:1986un}. This yields a structure of
divergences that departs from the flat space expression
(\ref{Kexp}) only by terms of order $a^4/r^4_0$ and higher. The
renormalized action is therefore finite for $d< 6$, with the
expected value. For $d=6$ the action is finite and nonvanishing,
but different for the two counterterms; and for $d>6$ the
divergences are not cancelled. This indicates that for $d>6$ the
spheroidal boundary deviates too strongly from the round sphere
(for which both counterterms $\cta$ and $\ctb$ were designed).
Presumably there exists another more sophisticated counterterm
which subtracts off all divergences for $d>6$ spheroidal
boundaries as well. We have not tried to construct such a
counterterm.

\subsubsection{Concluding Remarks}
The fact that flat space in spheroidal coordinates is assigned a
nonzero action in $d=6$ is at first surprising, but becomes less
so when we recall the analogous situation in AdS.  There we know
that the simplest choice of counterterms assign vanishing action
to AdS$_{d+1}$ in Poincar\'e coordinates, but finite action to
AdS$_{d+1}$ in global coordinates for $d$ even. This fact has a
nice interpretation in terms of the AdS/CFT correspondence: the
boundary of AdS in global coordinates is $S^{d-1} \times \RR$,
upon which the CFT can have a nonzero action due to the Casimir
effect. The ability of the counterterm prescription to assign
nonzero action to AdS in various coordinates is crucial to the
consistency of the AdS/CFT correspondence.  Although we do not
currently have access to a holographic description of AFS, we
should not be surprised that the action behaves in a way analogous
to AdS.

Since $\cta$ and $\ctb$ can lead to different finite terms in the
action, one can ask whether this implies that one, both, or
neither of them is in some sense ``correct''.  We believe that
both are valid expressions for the action, and should be thought
of as the results in  different renormalization schemes.  The
important criterion is that the counterterm ${\tilde S}$ can be
written in terms of intrinsic invariants of the boundary, and that
it removes the divergences of the action. In this sense, $\cta$
and $\ctb$ are both valid for boundaries of topology $S^{d-1}
\times \RR$, while only $\ctb$ is valid for the more general case
of $S^{n}\times \RR^{d-n}$. Future work may identify the most
general counterterm that subtracts the divergences of any
regularization of AFS. This would be interesting for several
purposes, including the general understanding of the asymptotic
symmetry group (the Spi group). However, from a practical point of
view, this development is unnecessary: our work shows that, when
the divergences of the action are cancelled for some counterterm,
the finite part has necessarily been correctly identified.

\vspace{0.2in} {\bf Acknowledgments:} We thank V. Balasubramanian for
collaboration in the initial stages of this work and for further discussions.
We also thank A. Ashtekar, R. Mann
and M. Taylor-Robinson for helpful discussions. This work was supported
by DOE Grant no. DE FG0290ER40560 (PK, FL),
by NSF Grant no. PHY 9600697 (PK),
by a Robert R. McCormick fellowship (FL). 
RS thanks the Enrico Fermi Institute and the Particle Theory Group of 
the University of Chicago for support. 
FL also thanks the Theory Groups at Harvard University and
at the University of Michigan in Ann Arbor for hospitality.


\end{document}